\documentclass[%
reprint,
superscriptaddress,
nofootinbib,
amsmath,amssymb,
aps,
longbibliography,
]{revtex4-1}

\usepackage{graphicx}
\usepackage{dcolumn}
\usepackage{bm}
\usepackage[mathlines]{lineno}

\begin{document}
\preprint{APS/123-QED}

\title{Transport property analysis method for thermoelectric materials: \\ material quality factor and the effective mass model}
\author{Stephen Dongmin Kang}
\affiliation{Department of Applied Physics and Materials Science, California Institute of Technology, CA 91125, USA}
\affiliation{Department of Materials Science and Engineering, Northwestern University, IL 60208, USA}
\author{G. Jeffrey Snyder}
\affiliation{Department of Materials Science and Engineering, Northwestern University, IL 60208, USA}



\begin{abstract}
Thermoelectric semiconducting materials are often evaluated by their figure-of-merit, $zT$. However, by using $zT$ as the metric for showing improvements, it is not immediately clear whether the improvement is from an enhancement of the inherent material property or from optimization of the carrier concentration. Here, we review the quality factor approach which allows one to separate these two contributions even without Hall measurements. We introduce practical methods that can be used without numerical integration. We discuss the underlying  effective mass model behind this method and show how it can be further advanced to study complex band structures using the Seebeck effective mass. We thereby dispel the common misconception that the usefulness of effective band models is limited to single parabolic band materials.
\end{abstract}

\maketitle

\section{The effective mass model}

In semiconducting band conductors, charge transport properties of interest are typically governed by the states near the band edge. Because the dispersion relation \textit{at the band edge} is typically parabolic ($E=\hbar^2 k^2/m^*$), it is often helpful to use an effective mass ($m^*$) model to characterize experimentally measured transport data.  The general approach is to consider the electronic structure of the majority carriers, whether holes or electrons, to be described by an effective mass $m^*$ that is independent of temperature and doping level. This approach puts our primary interest on data where transport contribution from minority carriers is not significant.  

We introduce in the next section a simple and accessible method -- the quality factor approach -- to analyze transport data using an effective mass model without the need for performing numerical integration of the Fermi function or even explicitly determining the effective mass. Even the simplest application of this model by using only thermopower ($|S|$), electrical conductivity ($\sigma$), and thermal conductivity ($\kappa$) measurements allows one to predict the maximum $zT=TS^2\sigma/\kappa$ that would be expected from optimizing the doping.  With the further use of Hall measurements to extract a value for $m^*$, the effective mass model makes it easy to identify complexities in the band structure and compare to theory.

\begin{figure}[t]
	\centering
	\includegraphics[width=0.3\textwidth]{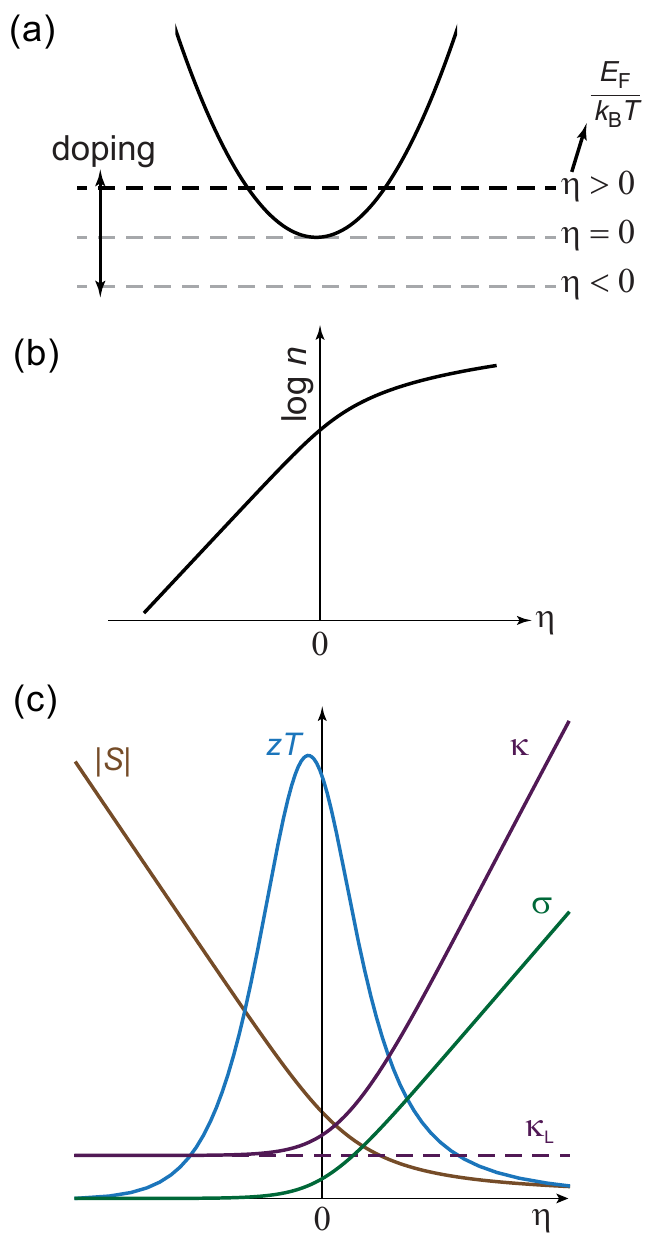}
	\caption{Reduced chemical potential $\eta$ for transport modeling. \textbf{(a)} $\eta$ is defined as the Fermi level $E_\mathrm{F}$ measured from the band edge, divided by $k_\mathrm{B}T$. Changing the carrier concentration by doping is equivalent to adjusting $\eta$. \textbf{(b)} Carrier concentration is a monotonically increasing function of $\eta$. \textbf{(c)} Thermopower ($|S|$) decreases with $\eta$ while electrical and thermal conductivities increase, making $zT$ highest at an optimum $\eta$.
	}	
	\label{fig:BrelatedSchematics}
\end{figure}

\section{Material Quality Factor Analysis Using Only $S$, $\sigma$, and $\kappa$}

Thermoelectric materials research typically aims to identify good thermoelectric materials and optimize their properties so that they can achieve the best possible $zT$. Since the $zT$ of a material peaks at an optimum carrier concentration (Fig.\ref{fig:BrelatedSchematics}), measuring $zT$ of one sample does not immediately provide an idea of the ultimate potential of a given material for thermoelectrics; a material initially measured with $zT < 0.1$ might end up with $zT > 1$ after tuning the carrier concentration.

Given the conductivity, Seebeck coefficient, and thermal conductivity of a single sample at an arbitrary doping level, what would be the best guess for its highest $zT$ expected after optimizing its carrier concentration?  Should one increase or decrease the amount of free carriers? These questions can be answered even without a Hall or mobility measurement. The quality factor analysis, \cite{Chasmar1959, Wang2013} based on an effective mass model, is devised to aid in the search for good thermoelectric materials by providing a convenient means for finding these answers.

The essence of the approach is to treat $zT$ as a function of two independent variables: the reduced Fermi level (reduced electron chemical potential) $\eta = E_\mathrm{F}/k_\mathrm{B}T$ (Fig.\ref{fig:BrelatedSchematics}a), and the ``material quality factor $B$.''  The former is a function of doping and temperature, and can be extracted from the Seebeck coefficient. In the steps described below, it is, in fact, not necessary to directly calculate a value of $\eta$. The latter is a material property largely independent of doping (though still dependent on temperature) given by \cite{Kang2017}:
\begin{equation}\label{eq:BfactorDef}
B = \left(\frac{k_\mathrm{B}}{e}\right)^2\frac{\sigma_\mathrm{E_0}}{\kappa_\mathrm{L}}T.
\end{equation}
Here, $k_\mathrm{B}$ is the Boltzmann constant, $\kappa_\mathrm{L}$ is lattice thermal conductivity and $\sigma_\mathrm{E_0}$ is a transport coefficient with units of conductivity that characterizes how well a material conducts electricity for a given $\eta$ (\textit{i.e.}, at a given carrier concentration). The material quality factor effectively removes all dependences on $\eta$ (\textit{i.e.}, on carrier concentration), and retains only the \emph{inherent} material properties that determine $zT$. This approach is successful because both $m^*$ (which is encompassed in $\sigma_\mathrm{E_0}$)  and $\kappa_\mathrm{L}$ remain relatively constant for the range of $\eta$ values that is experimentally tested by changing the carrier concentration (\textit{e.g.} doping, Fig.\ref{fig:BrelatedSchematics}b).

The thermopower $|S|$ at any temperature or doping concentration is best described as a function of only $\eta$ \cite{Fistul1969, May2012}: $|S(\eta)|$ (schematically shown in Fig.\ref{fig:BrelatedSchematics}c). Thermopower is merely an indicator of $\eta$ or $E_\mathrm{F}$. High $|S|$ does \textit{not} necessarily indicate a high quality thermoelectric material, nor does it directly determine the quality factor $B$. For semiconductors that can be doped, $S$ indicates the doping level which depends on defects and impurities; making tabulated values of $S$ for ``pure'' semiconductors or insulators is virtually meaningless. To optimize the $zT$ of a material, $\eta$, and thus $S$, must be tuned to an optimal value via doping (Fig.\ref{fig:BrelatedSchematics}c). In the method presented here, we simply use $S$ as a direct indicator of the doping level -- there is no need to calculate $\eta$.

\begin{figure}[tb]
	\centering
	\includegraphics[width=0.3\textwidth]{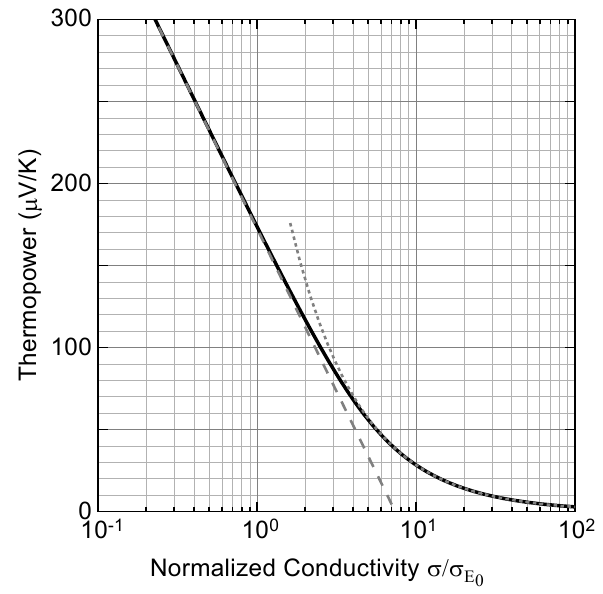}
	\caption{Determining $\sigma_\mathrm{E_0}$ from a $S$-$\sigma$ pair. Thermopower ($|S|$) determines the $\sigma$/$\sigma_\mathrm{E_0}$ of a sample, which allows one to determine $\sigma_\mathrm{E_0}$ from a pair of measured $S$ and $\sigma$. The analytical relations for the high $S$ limit (Eq.\ref{eq:SeebCondApproxNonDegen}) (dashed line) and low $S$ limit (Eq.\ref{eq:SeebCondApproxDegen}) (dotted line) are also shown together.
	}	
	\label{fig:Sigma_E0_determination}
\end{figure}

Conductivity, on the other hand, depends on $\sigma_\mathrm{E_0}$ as well as $\eta$:
\begin{equation}\label{eq:ConductivityWithSigmaE0}
\sigma = \sigma_\mathrm{E_0} \cdot \ln(1+e^\eta).
\end{equation}
Here, the $\eta$ term describes the increase in charge carriers as the Fermi level is increased. It is seen that $\sigma_\mathrm{E_0}$ describes the conductive ``quality'' of charge carriers in the material (magnitude of conductivity for a given $\eta$). Typically, $\sigma_\mathrm{E_0}$ is broken down to $m^*$ and the mobility parameter $\mu_0$ (determination of each requires a Hall measurement), but this decomposition is not always necessary for a basic use of the quality factor analysis. $\sigma_\mathrm{E_0}$ can be estimated from a pair of $S$ and $\sigma$ measurements on the same sample. As shown in Fig.\ref{fig:Sigma_E0_determination}, $S$ vs. $\sigma/\sigma_\mathrm{E_0}$ follows a universal curve; \textit{i.e.} one can graphically find the $\sigma/\sigma_\mathrm{E_0}$ that corresponds to the measured $S$ to find $\sigma_\mathrm{E_0}$. Alternatively, one can use the analytical expressions in the limits when is $|S|$ is large (within $5\%$ when $|S|>120$~$\mu$V/K):
\begin{equation}\label{eq:SeebCondApproxNonDegen}
\sigma_\mathrm{E_0}=\sigma\cdot\exp\left[\frac{|S|}{k_\mathrm{B} / e}-2\right],
\end{equation}
or small (within $5\%$ when $|S|<75$~$\mu$V/K):
\begin{equation}\label{eq:SeebCondApproxDegen}
\sigma_\mathrm{E_0} = \sigma \cdot \frac{3}{\pi^2}\frac{|S|}{k_\mathrm{B} / e}.
\end{equation}
The graphical method is better for intermediate $|S|$ values.

Low lattice thermal conductivity, $\kappa_\mathrm{L}$, is also a relevant descriptor for a good thermoelectric material because $\kappa_\mathrm{L}$ is typically independent of $\eta$. $\kappa_\mathrm{L}$ is obtained by subtracting, from the measured $\kappa$, the electronic portion ($\kappa_\mathrm{e}$) which is $\eta$ dependent:
\begin{equation}\label{eq:KappaDecomposed}
\kappa_\mathrm{L} = \kappa - \kappa_\mathrm{e} = \kappa - L \sigma T.
\end{equation}
Here, the Lorenz number $L$, defined by $\kappa_\mathrm{e} = L \sigma T$, is also a function of only $\eta$ (like $S(\eta)$) \cite{Fistul1969, May2012}: $L(\eta)$. Keeping in mind that $S$ is the experimental indicator of $\eta$, the value of $L$ at a given temperature can be approximated using measurements of $S$ using: \cite{Kim2015}
\begin{equation}\label{eq:LorenzApproximate}
L~[10^{-8}~\mathrm{W}\Omega/K^2 ] \approx 1.5 + \exp\left(-\frac{|S|}{116~\mu \mathrm{V/K}}\right).
\end{equation}

To see how the definition of $B$ in Eq.\ref{eq:BfactorDef} is justified, we can now separate the $\eta$-dependent terms from $zT$:
\begin{equation}
\begin{split}
zT &= \frac{S^2 \sigma T}{\kappa_\mathrm{L} + \kappa_\mathrm{e}} =  \frac{S^2}{\frac{\kappa_\mathrm{L}}{\sigma T} + L}\\ &= \frac{S^2(\eta)}{\frac{\kappa_\mathrm{L}}{T\sigma_\mathrm{E_0} \cdot \ln(1+e^\eta)}+L(\eta)} \\&= \frac{S^2(\eta)}{\frac{(k_\mathrm{B}/e)^2}{B \ln(1+e^\eta)}+L(\eta)},
\end{split}
\end{equation}
where $B$ combines all the $\eta$-independent material parameters, giving the definition of the \textit{dimensionless material quality factor} in Eq.\ref{eq:BfactorDef}. The natural unit of the Lorenz number $(k_\mathrm{B}/e)^2$ was multiplied in the term containing $1/B$ to make $B$ dimensionless for convenience (some authors \cite{May2012} use $\beta=B/ (k_\mathrm{B}/e)^2$).

\begin{figure}[tb]
	\centering
	\includegraphics[width=0.3\textwidth]{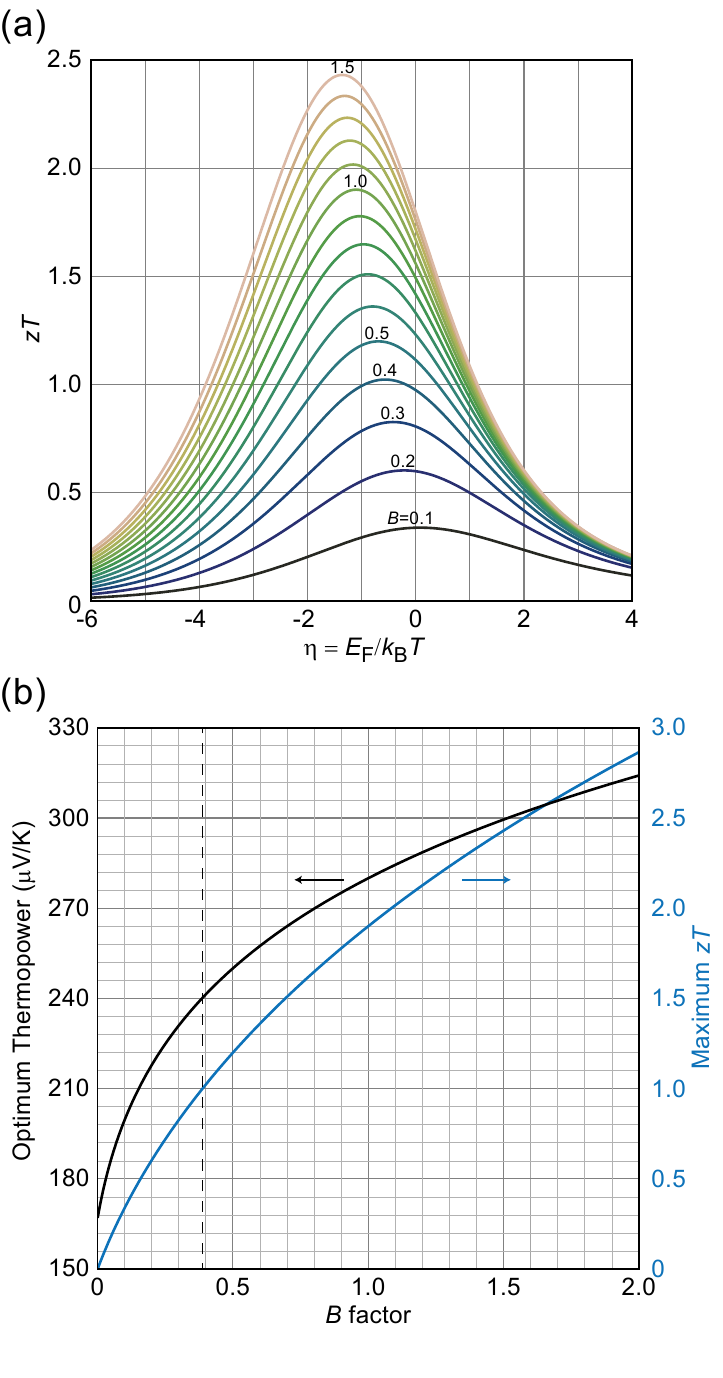}
	\caption{
		Material quality factor analysis. \textbf{(a)} The $zT$~vs.~$\eta$ relation is determined by the material quality factor $B$, making maximum $zT$ and optimum $\eta$ a function of $B$. \textbf{(b)} Thermopower at optimum $\eta$ as a function of $B$ (black line, left axis), which can be used as a guide for optimization. Maximum $zT$ is also plotted together (blue line, right axis). The vertical dashed line indicates when maximum $zT=1$, corresponding to $B\approx0.4$ which serves as a convenient reference value for a good thermoelectric material.
	}	
	\label{fig:BAnalysis}
\end{figure}

This quality factor $B$ completely determines the $zT$ vs. $\eta$ curve (Fig.\ref{fig:BAnalysis}a) for a given material at a given temperature. Therefore, $B$ is a good descriptor to estimate the maximum $zT$ achievable from a material when the carrier concentration (and, thus $\eta$) is optimized; $B$ also determines the optimum level of doping (Fig.\ref{fig:BAnalysis}b). Practically, tuning towards the optimum is most easily done by looking at the optimum thermopower that is expected from a given $B$. For example, if $B=0.4$ and $S=50$~$\mu$V/K was obtained from a sample at a given temperature, one can expect to reach $zT>1$ by decreasing the carrier concentration until $S=240$~$\mu$V/K.

Given $S$, $\sigma$, and $\kappa$ of a single sample at a given temperature, one can estimate $\sigma_\mathrm{E_0}$ and $\kappa_\mathrm{L}$, which allows the estimation of $B$ at that temperature using Eq.\ref{eq:BfactorDef}. The approximate methods (Eqs.\ref{eq:LorenzApproximate} and \ref{eq:SeebCondApproxNonDegen}-\ref{eq:SeebCondApproxDegen}) described above make this estimation quick and easy. The full calculation is also straightforward, but requires numerical integration and root finding using the expressions of $|S(\eta)|$ and $L(\eta)$ which can be found in Ref.[\onlinecite{May2012}]. The first step is to estimate $\eta$ from the measured $S$ by numerical solving. Then, one could use $\eta$ and the measured conductivity to estimate $\sigma_\mathrm{E_0}$ using Eq.\ref{eq:ConductivityWithSigmaE0}. Lastly, to estimate the lattice thermal conductivity $\kappa_\mathrm{L}$ from measured $\kappa$ (Eq.\ref{eq:KappaDecomposed}), one can calculate $L(\eta)$ using the $\eta$ estimated from $S$.

The application of the $B$-factor $zT$ analysis is simple once $\sigma_\mathrm{E_0}$ and $\kappa_\mathrm{L}$ is determined. One can calculate $B$ from Eq.\ref{eq:BfactorDef}. Then Fig.\ref{fig:BAnalysis}b can be used to find the maximum $zT$ and optimum thermopower.

The approach of separating $\eta$ dependency (or doping dependency) from intrinsic material parameters can be extended to atypical cases, such as conducting polymers \cite{Kang2017}, where transport behaves in a different way (different energy dependency of transport and different scale of $B$) than found in inorganic crystalline materials. 

\section{Bipolar Effects}

A particular material is well described by a single $B$ as long as the carriers in the $\eta$ range of interest are well characterized by a single $m^*$. The most common situation in which a single $m^*$ does \textit{not} suffice is when there is non-negligible contribution from minority carriers. This bipolar transport happens in all semiconductors at high temperatures (\textit{i.e.}, when $k_\mathrm{B}T$ becomes comparable to $\approx E_\mathrm{g}/4$, where $E_\mathrm{g}$ is the band gap). The onset of bipolar conduction is best identified from the thermopower showing a flattening or rollover with increasing temperature (\textit{i.e.} diminishing slope in $|S|$~vs.~$T$). It is possible to estimate $B$ of the majority carriers in the bipolar region by extrapolating $\sigma_\mathrm{E_0}$~vs.~$T$ from the non-bipolar region. In the most common case of acoustic-phonon scattering, $\sigma_\mathrm{E_0}$ is nearly constant with respect to $T$. At a temperature where bipolar conduction dominates, one would realize that the optimum $S$ required for maximum $zT$, as evaluated from the $B$ of majority carriers, is not obtainable at that temperature due to the canceling contribution of minority carriers. The maximum thermopower obtainable ($|S_\text{max}(T_\text{max})|$) is related to the band gap ($E_\mathrm{g} \approx 2e|S_\text{max}|T_\text{max}$) \cite{Gibbs2015}, demonstrating how the maximum $zT$ becomes band-gap limited. An example calculation can be found in Ref.[\onlinecite{Kang2017a}], where the effective overall $B$ is smaller than that of the majority carriers due to bipolar contribution.

A higher peak $zT$ value is obtainable from a larger band gap for a given $\sigma_\mathrm{E_0}$ of the majority carriers. The temperature at which the peak $zT$ is found increases with a larger band gap, leading to a higher $zT$. This principle motivates to tune the band gap (\textit{e.g.} by alloying) either to increase the peak $zT$ or to shift the peak $zT$ temperature. Because the band gap and $\sigma_\mathrm{E_0}$ are often not independent to each other and material stability limits the maximum temperature of a material, the optimum band gap tends to depend on the material and application.

It is worth to note that the bipolar effect from a given band gap can be suppressed if the majority carriers have a higher $\sigma_\mathrm{E_0}$ than minority carriers. In this sense, materials with highly contrasted conduction and valence band structures have a larger effective gap when doped with its superior type of carriers.

\section{Effective $m^*$ for studying complex electronic structures}

Evaluating $\sigma_\mathrm{E_0}$ (but not $m^*$) was sufficient for the quality factor analysis; assessment of $m^*$ offers a further step through which one can study the band structure of materials using transport measurements. 

The equations used so far (Eqs.\ref{eq:ConductivityWithSigmaE0}-\ref{eq:SeebCondApproxDegen}) are from a model of free carriers (\textit{i.e.} parabolic dispersion) being scattered by acoustic phonons \cite{Fistul1969} and is sometimes referred to as the single parabolic band model; however, the use of these equations does not necessarily require a single parabolic band assumption. Even for complicated band structures that are non-single or significantly non-parabolic, we can build upon the same approach to characterize the free-carrier equivalent, an effective $m^*$ that can change with temperature and energy. Then, one can relate certain band complexities to particular trends in $m^*$.

For this purpose, we can break down $\sigma_\mathrm{E_0}$ in terms of $m^*$. In anticipation that, in the case of non-simple band structures, $m^*$ will be differently determined depending on how it is assessed, we will distinguish $m^*$'s with a subscript. In band conductors, $\sigma_\mathrm{E_0}$ is \cite{Kang2017}:
\begin{equation}\label{eq:TransportCoefficientInBandTerms}
\sigma_\mathrm{E_0} = \frac{{8\pi e{{\left( {2{m_e}k_\mathrm{B} T} \right)}^{3/2}}}}{{3{h^3}}} \cdot {\mu _0}{\left(\frac{{{m^*_\mathrm{S}}}}{{{m_e}}}\right)^{3/2}}.
\end{equation}
Here, $\mu_0=e \tau_0 /m^*_\mathrm{I}$ is a mobility parameter, where $\tau_0$ describes the relaxation time of carriers through $\tau = \tau_0\cdot (E/k_\mathrm{B}T)^{-1/2}$ and $m^*_\mathrm{I}$ is the inertial effective mass. $m^*_\mathrm{S}$ is the Seebeck effective mass and $m_e$ is the mass of an electron.

The quantity $\mu_\text{w} = \mu_0({m^*_\mathrm{S}}/{m_e})^{3/2}$ is called the weighted mobility and is directly proportional to $\sigma_\mathrm{E_0}$ for a given $T$. Some authors \cite{Wang2013} use the non-degenerate limit drift mobility ($\mu_{\text{cl}} = 4/3\sqrt{\pi} \cdot \mu_0$) to define $\mu_0$ (and thus, $\mu_\text{w}$)\footnote{Inconsistently interchanging $\mu_0$ and $\mu_{\text{cl}}$ could lead to errors in $B$ by a factor of $3\sqrt{\pi}/4\approx1.33$ (\textit{e.g.} Fig.1 in Ref.\cite{Pei2012}).}.

The Seebeck effective mass describes the number of states for a given reduced chemical potential $\eta$, where $\eta$ is evaluated using $|S|$ and the number of states at that $\eta$ is evaluated using a Hall measurement. $m^*_\mathrm{S}$ can be calculated within $2\%$ by using the following equations. When $|S|>75$~$\mu$V/K:

\begin{equation}\label{eq:SeebeckMass1}
m^*_\mathrm{S} \approx \frac{h^2}{2k_\mathrm{B}T} \left\{\frac{3n_\mathrm{H}}{16\sqrt{\pi}} \left(\exp\left[\frac{|S|}{(k_\mathrm{B}/e)}-2 \right] -0.17\right ) \right\}^{2/3}.
\end{equation}
and, when $|S|<75$~$\mu$V/K:
\begin{equation}\label{eq:SeebeckMass2}
m^*_\mathrm{S} \approx \frac{3h^2}{8\pi^2k_\mathrm{B}T} \frac{|S|}{(k_\mathrm{B}/e)} \left(\frac{3n_\mathrm{H}}{\pi}\right)^{2/3}.
\end{equation}
Here, $n_\mathrm{H}$ is the Hall carrier concentration. A heavier $m^*_\mathrm{S}$ gives higher $|S|$ for a given $n_\mathrm{H}$ (alternatively, a higher $n_\mathrm{H}$ for a given $|S|$). A plot of $|S|$ with respect to $n_\mathrm{H}$ is called a ``Pisarenko plot'', by which one can determine $m^*_\mathrm{S}$ from a set of data points.

The inertial mass is not easily separable from the relaxation time, and we thus keep it in the form of $\mu_0=e \tau_0 /m^*_\mathrm{I}$. We can nevertheless understand how band structure impacts $\mu_0$ using the deformation potential model by Bardeen and Shockley \cite{Bardeen1950}:
\begin{equation} \label{eq:mu0_acousticphononscattering}
\mu_0 = \frac{\pi e\hbar^4C_l}{\sqrt{2} m^*_\mathrm{I}m_\mathrm{b}^*{}^{3/2}(k_\mathrm{B}T)^{3/2}\Xi^2}.
\end{equation}
Here, $m_\mathrm{b}^*$ is the effective mass that describes the density-of-states of an individual Fermi-surface pocket. $\Xi$ is the deformation potential and $C_l$ is the longitudinal elastic constant.

The combination of Eqs.\ref{eq:SeebeckMass1}-\ref{eq:SeebeckMass2} and \ref{eq:mu0_acousticphononscattering} helps one understand what type of band is good for thermoelectrics: $\mu_\text{w} \propto ({m_S^*/m_\mathrm{b}^*})^{3/2} / m^*_\mathrm{I}$. Suppose that symmetry provides $N_\mathrm{V}$ multiple bands with the same dispersion (multi-valley degeneracy). Then $\mu_0$ of a single band is identical to the $\mu_0$ of all the multiple bands together. On the other hand, $m_S^*$ is larger than that of a single band by $N_\mathrm{V}^{2/3}$ because the density-of-states (and thus $n_\mathrm{H}$ in Eqs.\ref{eq:SeebeckMass1}-\ref{eq:SeebeckMass2}) is larger by a factor of $N_\mathrm{V}$. Overall, $\mu_\text{w}$ (and, thus $\sigma_\mathrm{E_0}$ and $B$) scales with $N_\mathrm{V}/m^*_\mathrm{I}$. Therefore, multi-valleys and lighter bands (small $m^*_\mathrm{I}$) are advantageous for thermoelectrics.

In general, when multiple bands contribute to transport, they are not necessarily identical or aligned; nevertheless, the trend of $m_S^*$ and $\mu_\text{w}$ both increasing simultaneously with advantageous band complexity remains similar \cite{Gibbs2017}, allowing one to relate transport measurements to understandings of the electronic structure. Therefore, it is best to keep track of both $m_S^*$ and $\mu_\text{w}$ when analyzing transport data.

An advanced example would be a case when two conduction bands have their band edges offset by a small amount on the order of a few $k_\mathrm{B}T$. When $E_\mathrm{F}$ is below the lower band, the upper band would not contribute significantly to transport. With doping, once $E_\mathrm{F}$ moves within a few $k_\mathrm{B}T$ to the edge of the upper band, both bands would start contributing. In experimental characterization, one would observe $m_S^*$ and $\mu_\text{w}$ both increasing at a threshold of $\eta$, where the threshold indicates how much the bands are offset from each other. Such a signature would be a strong motivation to further investigate the band structure using more specific methods such as optical absorption.

The usefulness of $m_S^*$, or any $m^*$ in general, comes from the fact that it is a convenient metric to characterize an electronic structure and so used to characterize diverse measurements such as the electronic specific heat, plasma frequency, as well as Seebeck coefficient. Mathematically, the procedure could be understood as a change of variables. While $E$, $k$, $\sigma_E$, $\tau$, or density-of-states change dramatically with experimental variables such as doping or temperature, the various $m^*$'s as defined though different measurements (Seebeck, specific heat, plasma frequency, etc) remain relatively constant and thus $m^*$'s are typically reported as results of such measurements. Just as $m^*$'s are reported rather than specific values of electronic specific heat (\textit{e.g.} heavy fermion metals) or optical absorption (plasma frequency measurements), it would be more useful to report $m_S^*$ than specific values of $S$ in many insulators and semiconductors. All of these effective masses are expected to change somewhat with doping, temperature and even alloying and structural modification. In fact, observing and quantitatively characterizing how $m^*$ changes might be the best way to identify changes in parabolicity or multiple band effects \cite{Wang2014, Tang2015, Kim2017}. In this way, the effective $m^*$ approach does not simply assume, or impose an approximation of, a single parabolic band, but rather provides a helpful means to characterize data and identify deviations from single or parabolic electronic structures.

\bibliography{BfactorRefs.bib}

\end{document}